\documentclass[aps,preprint,prd,,nofootinbib]{revtex4-1}
\usepackage{amssymb}
\usepackage{amsmath}
\usepackage{graphicx}
\usepackage{fancyhdr}
\def\br{\begin{eqnarray}}
\def\er{\end{eqnarray}}
\def\be{\begin{equation}}
\def\ee{\end{equation}}

\def\({\left(}
\def\){\right)}

\newcommand \beq { \begin{eqnarray} }
\newcommand \eeq { \end{eqnarray} }
\newcommand \beqq { \begin{equation} }
\newcommand \eeqq { \end{equation} }
\begin{document}

\title{The minimal 3-3-1 model with only two Higgs triplets}

\author{J. G. Ferreira Jr, P. R. D. Pinheiro,  C. A. de S. Pires, P. S. Rodrigues da Silva}
\email{prdpinheiro@fisica.ufpb.br,
cpires@fisica.ufpb.br, psilva@fisica.ufpb.br}
\affiliation{{ Departamento de
F\'{\i}sica, Universidade Federal da Para\'\i ba, Caixa Postal 5008, 58051-970,
Jo\~ao Pessoa, PB, Brasil}}



\date{\today}

\begin{abstract}
The simplest non-abelian gauge extension of the electroweak standard model, the $SU(3)_c\otimes SU(3)_L\otimes U(1)_N$, known as 3-3-1 model, has a minimal version which demands the least possible fermionic content to account for the whole established phenomenology for the well known particles and interactions. Nevertheless, in its original form the minimal 3-3-1 model was proposed with a set of three scalar triplets and one  sextet in order to yield the spontaneous breaking of the gauge symmetry and generate the observed fermion masses. Such a huge scalar sector turns the task of clearly identifying the physical scalar spectrum a clumsy labor. It  not only adds an obstacle for the development of its phenomenology but implies a scalar potential plagued with new free coupling constants. In this work we show that the framework of the minimal 3-3-1 model can be built with only two scalar triplets, but still triggering the desired pattern of spontaneous symmetry breaking and generating the correct fermion masses. We present the exact physical spectrum and also show all the interactions involving the scalars, obtaining a neat minimal 3-3-1 model far more suited for phenomenological studies at the current Large Hadron Collider.
\\
\end{abstract}

\maketitle
\section{Introduction}

One of the main goals of the Large Hadron Collider~(LHC) is  to find the Higgs boson and  probe new physics at TeV scale. This opens a timely window to review the scalar sector of extensions of the electroweak standard model~(SM) at the particular range  of energy covered by the LHC.
Here we specialize in an simple non-abelian gauge extension of SM, the $SU(3)_c\otimes SU(3)_L\otimes U(1)_N$, the so-called 3-3-1 model, whose characteristic scale lies at TeV scale. Among the available versions,  the minimal 3-3-1 model~\cite{Vicente} may be considered the most phenomenologically interesting one because of its reduced leptonic content and the presence of bileptons (fields carrying two units of lepton number): a singly and a doubly charged gauge bosons, $V^\pm$ and $U^{\pm\pm}$, and new quarks with exotic electric charges $\frac{5}{3}e$ and  $\frac{4}{3}e$, besides scalar bileptons. 
In a lepton number conserving framework these new particles are constrained to involve a pair of sibling leptons in their decay branch, posing a singular signature to be searched for in colliders.
From the theoretical side,  the model  predicts the existence of three fermion generations, solving the so called family replication problem, explains why the  electric charge is quantized\cite{ecq}, and allows in its structure the realization of the Peccei-Quinn symmetry which may provide a solution to  the strong-CP problem\cite{PQ}. These are some of the features that turns the minimal 3-3-1 model an excellent proposal of physics beyond the SM, justifying further efforts to make it phenomenologically more appealing. In this way, we should look for some room for improvement without jeopardizing any of the nice features of the model.

It happens that one of the unpleasant parts of the model is exactly the one that offers the opportune window to ameliorate its framework. It concerns its huge scalar sector\cite{foot}, composed by three triplets and one sextet\footnote{If one intends to explain the correct mass for all leptons, including neutrinos, an extra sextet should be added~\cite{2sextetos} and the number of degrees of freedom are really much bigger.}, totalizing thirty degrees of freedom.
The spontaneous breaking of the 3-3-1 model gauge symmetry requires eight degrees of freedom in the form of Goldstones, meaning that after the breaking there remains twenty two degrees of freedom in the spectrum as physical scalars.
It would be understandable that some prejudice could be raised against this model relying on such a prolific scalar sector, since we are still waiting for the first fundamental scalar to be detected. 
We mention that besides having a large amount of additional scalar degrees of freedom compared to the SM, the gauge symmetry also allows for a scalar potential that implies several unknown coupling constants among these scalars. Finally, such a scalar potential leads to intricate mass matrices that are very awkward to deal with\cite{tonasse}. In other words, we have to rely upon several suppositions concerning the couplings and the energy scales of the model in order to obtain a simpler and treatable physical scalar spectrum (mass eigenstates). However, as the LHC is focused on the Higgs boson search, it is imperative that a testable model have a neat and manageable scalar spectrum, with the lowest number of free parameters and all the couplings well defined, so as to better contrast the LHC findings with the model predictions.
Then, if we are able to reduce the number of scalars in this model it would be a huge gain, in number and simplicity.

In this work we argue that such an approach is possible for the minimal 3-3-1 model. Next we present the reduced minimal 3-3-1 (RM331) model and implement this scheme by showing that only two scalar triplets are sufficient to break correctly the $SU(3)_C \times SU(3)_L \times U(1)_N$ symmetry to the $SU(3)_C \times U(1)_{QED}$ and generates the correct masses of all fermions including neutrinos.

\section{The reduced minimal 3-3-1 model}
As in the original version~\cite{Vicente}, each family of leptons comes in a triplet representation of $SU(3)_L$,
\begin{equation}
\begin{array}{cc}
f_L=\left(\begin{array}{c}
\nu_l \\  l\\ l^c
\end{array}\right)_L \sim (1,3,0),
\end{array}
\label{leptonscontent}
\end{equation}
where $ l=e,\mu ,\tau$. The numbers in parenthesis represent the fields' transformation properties under the gauge group $SU(3)_C \times SU(3)_L \times U(1)_N$.

In the quark sector, anomaly cancellation requires that  one generation comes in a $SU(3)_L$ triplet and the other two come in anti-triplet representation  with the following content,
\begin{eqnarray}
&&
\begin{array}{c}
 Q_{1L}=\left(\begin{array}{cc}
 u_1 \\ d_1 \\ J_1
\end{array}\right)_L \sim (3,3,+\frac{2}{3})
\end{array}\,\,,\,\,
\begin{array}{c}
Q_{iL}=\left(\begin{array}{cc}
 d_i \\ -u_i \\ J_i
\end{array}\right)_L \sim (3,3^*,-\frac{1}{3}),
\end{array}
\nonumber \\
&&
\begin{array}{ccc}
u_{iR} \sim(3,1,+\frac{2}{3}); & d_{iR}
\sim(3,1,-\frac{1}{3}); & J_{iR}
\sim(3,1,-\frac{4}{3}),
\end{array}
\nonumber \\
&&
\begin{array}{ccc}
u_{1R} \sim(3,1,+\frac{2}{3}); & d_{1R}
\sim(3,1,-\frac{1}{3}); & J_{1R}
\sim(3,1,+\frac{5}{3}),
\end{array}
\end{eqnarray}
\label{quarkcontent}
where $i=2,3$.

We now introduce the main point of our proposal. Instead of three triplets and one sextet of scalars as in the original version, we propose  a RM331 model with only  two scalar triplets\footnote{It is important to stress that our proposal would work also for the alternative choice of keeping the scalar triplet $\eta = (\eta^0,\, \eta_1^-\,, \eta_2^+)\sim (1,\,3,\,0)$ instead of $\rho$. However, in this case neutrino masses would arise through the dimension-five effective operator, $\frac{h}{\Lambda}(\bar L^C \eta^*) (\eta^\dagger L)$, leading to a neutrino mass scale two orders of magnitude higher than the dimension-7 operator we are going to obtain in this work. Thus our choice for $\rho$ is based  in  obtaining the higher effective operator for neutrino masses, demanding less fine-tuning in the respective coupling.},
\begin{equation}
\begin{array}{cccc}
\rho =\left(\begin{array}{c}
\rho^+ \\ \rho^0 \\ \rho^{++}
\end{array}\right) \sim (1,3,1);&
\chi =\left(\begin{array}{c}
\chi^- \\ \chi^{--} \\ \chi^0\end{array}\right) \sim (1,3,-1).
\end{array}
\label{scalarcontent}
\end{equation}

Before going into technical details, we first argue that such short scalar content is sufficient to engender the correct pattern of gauge symmetry breaking and generate masses for all fermions including neutrinos. 

Let us first analyze the spontaneous symmetry breaking sequence. It is easy to see that when $\chi^0$ develops a non-trivial vacuum  expectation value (VEV), $v_\chi$, the $SU(3)_C \times SU(3)_L \times U(1)_N$ symmetry breaks spontaneously to the $SU(3)_C \times SU(2)_L \times U(1)_Y$, which in turn breaks to $SU(3)_C \times U(1)_{QED}$, when $\rho^0$ develops a VEV, $v_\rho$. 

Although the breaking occurs as desired, we have to check that there are enough degrees of freedom  for supporting this spontaneous symmetry breaking pattern. To illustrate that this is the case, we recall that the charged scalars  $\chi^{\pm}$ are the only ones to carry two units of lepton number (bileptons)~\cite{Vicente} meaning that they are the Goldstone bosons absorbed by the $V^{\pm}$ gauge bosons (also bileptons), while $\rho^{\pm}$ are the unique singly charged scalars that do not carry lepton number, thus they should be the Goldstones eaten by the standard $W^{\pm}$. The doubly charged scalars,  $\rho^{\pm \pm}$  and $\chi^{\pm \pm}$, are both bileptons, and one linear combination of them should give the longitudinal component of the vector bilepton $U^{\pm \pm}$, while the orthogonal combination remains in the physical spectrum.  Regarding the neutral sector,  $I_\rho$  is the  Goldstone boson eaten by $Z_\mu$   and $I_\chi$ is the  Goldstone eaten by $Z^{\prime}_\mu$. So, from this short analysis, we conclude that the two scalar triplets possess the right number of degrees of freedom to engender the expected pattern of spontaneous symmetry breaking in the  RM331 model. As a consequence, from the initial 12 scalar degrees of freedom, only four of them survive in the physical spectrum, two doubly charged, $h^{\pm \pm}$, and two neutral ones, $h^1$  and $h_2$. One of the neutral scalars, the lightest, is going to be identified with the Higgs boson. All this makes it clear that the RM331 model keeps the main ingredients to become a successful gauge extension of SM. Next we explore the scalar sector in a little more detail.

\section{Scalar sector}
Considering the scalar content in Eq.~(\ref{scalarcontent}) we can write down the most general renormalizable, gauge and Lorentz invariant scalar potential, 
\begin{eqnarray}
V(\chi,\rho)&=&\mu^2_1\rho^\dagger\rho+
\mu^2_2\chi^\dagger\chi+\lambda_1(\rho^\dagger\rho)^2+\lambda_2(\chi^\dagger\chi)^2
\nonumber \\ & &\mbox{}
+\lambda_3(\rho^\dagger\rho)(\chi^\dagger\chi)+\lambda_4(\rho^\dagger\chi)(\chi^\dagger\rho)\,,
\label{potential}
\end{eqnarray}
whose manifest simplicity turns the RM331 model a real gain as compared to the original minimal 3-3-1 model since the number of free parameters is reduced from, at least, thirteen to only six.

Let us expand  $\rho^0$ and $\chi^0$ around their VEV's in the usual way,
\begin{equation}
\rho^0 \,,\, \chi^0 \rightarrow \frac{1}{\sqrt{2}}(v_{\rho\,,\,\chi}+ R_{\rho\,,\,\chi} +iI_{\rho\,,\,\chi}).
\label{VEVs}
\end{equation}

On substituting this expansion  in the above potential we obtain the following set of minimum constraint equations,
\begin{eqnarray}
&&
 \mu^2_1+\lambda_1 v^2_\rho+\frac{\lambda_3 v^2_\chi}{2}=0,\nonumber \\
 &&
 \mu^2_2+\lambda_2 v^2_\chi+\frac{\lambda_3 v^2_\rho}{2}=0.
\end{eqnarray}

Let us first consider the  doubly charged scalars. Their mass matrix in the basis $(\chi^{++}\,,\,\rho^{++})$ is given by,
\begin{equation}
m^2_{++}=\frac{\lambda_4 v^2_\chi}{2}
\begin{pmatrix}
t^2\ & t  \\
t& 1 
\end{pmatrix},
\label{++massmatrix}
\end{equation}
where $t=\frac{v_\rho}{v_\chi}$. Diagonalizing this matrix we obtain the following squared mass eigenvalues,
\begin{equation}
m^2_{\tilde h^{++}}=0\,\,\,\,\,\,\,\mbox{and}\,\,\,\,\,\,m^2_{h^{++}}=\frac{\lambda_4}{2}(v^2_\chi + v^2_\rho),
\label{masscharged}
\end{equation}
where the corresponding eigenstates are,
\begin{eqnarray}
&&\left( 
\begin{tabular}{c}
$\tilde h^{++}$ \\ 
$h^{++} $
\end{tabular}
\ \right) = \left( 
\begin{tabular}{cc}
$c_ \alpha$ & -$s_ \alpha$\\ 
$s_\alpha$ & $c_\alpha$
\end{tabular}
\ \right) \left( 
\begin{tabular}{l}
$\chi^{++} $ \\ 
$\rho^{++} $
\end{tabular}
\right),
\label{chargedeigenvectors}
\end{eqnarray}
with 
\begin{equation}
c_\alpha = \frac{v_\chi}{\sqrt{v^2_\chi + v^2_\rho}}\,,\,s_\alpha=\frac{v_\rho}{\sqrt{v^2_\chi + v^2_\rho}}.
\label{chargedeiggenvectors}
\end{equation}
It is easy to see that when $v_\chi >> v_\rho$ we have that $\tilde h^{++} \approx \chi^{++}$ and $h^{++}\approx \rho^{++}$. Notice that $\tilde h^{\pm \pm}$ are the Goldstones eaten by the gauge bosons $U^{\pm \pm}$, while $h^{\pm \pm}$ remains as a  physical scalar in the spectrum. 

Regarding the neutral scalars,  their mass matrix takes the following form in  the basis $(R_\chi\,,\,R_\rho)$,
\begin{equation}
m^2_0=\frac{v^2_\chi}{2}
\begin{pmatrix}
2\lambda_2\ & \lambda_3 t  \\
\lambda_ 3 t& 2\lambda_1 t^2 
\end{pmatrix}.
\label{neutralmassmatrix}
\end{equation}
On diagonalizing this mass matrix, we obtain the following eigenvalues,
\begin{equation}
m^2_{h^1}=(\lambda_1 -\frac{\lambda^2_3}{4\lambda_2})v^2_\rho\,\,\,,\,\,\,m^2_{h_2}=\lambda_2v^2_\chi +\frac{\lambda^2_3}{4\lambda_2}v^2_\rho,
\label{neutralmass}
\end{equation}
with their corresponding neutral physical eigenstates,
\begin{eqnarray}
h_1=c_\beta R_\rho-s_\beta R_\chi \,\,\,,\,\,\,
h_2=c_\beta R_\chi +s_\beta R_\rho \,,
\label{neutrales}
\end{eqnarray}
where $c_\beta \approx 1-\frac{\lambda^2_3}{8 \lambda^2_2}\frac{v^2_\rho}{v^2_\chi}$ and $
s_\beta \approx \frac{\lambda_3}{2\lambda_2}\frac{v_\rho}{v_\chi}$.

As for the pseudo-scalars, $I_\rho$  and $I_\chi$, they do not mix among themselves and are massless, meaning that they are Goldstone bosons eaten by the gauge bosons $Z_\mu$ and $Z^{\prime}_\mu$, respectively.

In the end of the day, the physical scalar spectrum of the RM331 model is composed by a doubly charged scalar $h^{++}$  and two neutral scalars $h_1$  and $h_2$. Since the lightest neutral field, $h_1$, is basically a $SU(2)_L$ component in the linear combination Eq.~(\ref{neutrales}), we identify it as the standard Higgs boson in this model.

\section{Gauge boson spectrum}
In order to obtain the expression for the masses of the massive gauge bosons of the model, we have to substitute the expansion of the Eq. (\ref{VEVs}) in the Lagrangian,
\begin{equation}
\mathcal{L=}\left( \mathcal{D}_{\mu }\chi \right) ^{\dagger }\left( \mathcal{%
D}^{\mu }\chi \right) +\left( \mathcal{D}_{\mu }\rho \right) ^{\dagger
}\left( \mathcal{D}^{\mu }\rho \right) ,
\label{derivada}
\end{equation}
where,
\begin{equation}
D_\mu = \partial_\mu -ig W^a_\mu \frac{\lambda ^a}{2}-ig_N N W^N_\mu,
\label{derivativecovariant}
\end{equation}
with  $a=1,...,8$ and $\lambda^a$ being the Gellmann matrices. 

On doing this, the  eigenstates of the charged gauge bosons and their respective masses are given by,
\begin{eqnarray}
&& 
W^{\pm}=\frac{W^1 \mp iW^2}{\sqrt{2}}\,\,\,\, \rightarrow \,\,\,\, M_{W^{\pm }}^{2} =\frac{g^{2}v_{\rho }^{2}}{4} , \nonumber \\
&&
V^{\pm}=\frac{W^4 \pm iW^5}{\sqrt{2}} \,\,\,\,\rightarrow  \,\,\,\,M_{V^{\pm }}^{2} =\frac{g^{2}v_{\chi }^{2}}{4} ,\nonumber \\
&&
 U^{\pm \pm}=\frac{W^6 \pm iW^7}{\sqrt{2}} \,\,\,\, \rightarrow \,\,\,\, M_{U^{\pm \pm }}^{2} =\frac{g^{2}\left( v_{\rho }^{2}+v_{\chi }^{2}\right) }{4},
\label{masseschargedbosons}
\end{eqnarray}

We call the attention to the interesting fact that a hierarchy arises among the charged gauge bosons. Namely, the mass expressions in Eq.~(\ref{masseschargedbosons}) provide  $M^2_U-M^2_V=M^2_W$, which is a direct consequence of the shortened scalar sector in the RM331 model.

From the three neutral gauge bosons, one is identified as the photon, $A_\mu$, which is massless, and  the other two are the standard $Z_\mu$  and a new $Z^{\prime}_\mu$, whose masses are given by,
\begin{equation}
m^2_Z=\frac{g^2}{4c^2_W}v^2_\rho \,\,\,\,,\,\,\,\,m^2_{Z^{\prime}}=\frac{g^2 c^2_W}{3(1-4 s^2_W)}v^2_\chi\,.
\label{neutralmassesbosons}
\end{equation}
We then write the massless gauge fields in terms of the mass eigenstates as follows,
\begin{eqnarray}
&&
W^N_\mu = -t_W\sqrt{h_W} Z_\mu + \sqrt{3}t_W Z^{\prime}_\mu + \sqrt{h_W}A_\mu, \nonumber \\
&&
W^8_\mu = \sqrt{3} t_W s_W Z_\mu +\frac{h_W}{c_W}Z^{\prime}_\mu -\sqrt{3}s_W A_\mu,\nonumber \\
&&
W^3_\mu = c_W Z_\mu + s_W A_\mu,
\label{neutraleigenstates}
\end{eqnarray}
where $c_W=\cos\theta_W$, $s_W=\sin\theta_W$, $t_W=\tan \theta_W$, $h_W = 1-4 s^2_W$, with $\theta_W$ being the Weinberg mixing angle.

Hence, we have four singly charged gauge bosons, the standard $W^{\pm}$  and the heavy one $V^{\pm}$, two doubly charged gauge bosons, $U^{\pm \pm}$, and three neutral gauge bosons, the photon $A_\mu$, the standard $Z_\mu$ and the heavy $Z^{\prime}_\mu$. Their trilinear and quartic interactions are the same as in Ref.~\cite{331gauge}, their neutral and charged current interactions with the fermions are given in the appendix~{\bf A} while their interactions with the physical scalars are presented in the appendix~{\bf B}.

\section{Fermion masses and proton decay}
The greatest impact of a shortened scalar content is that part of the fermion masses originates from Yukawa couplings and another part are due to effective operators. Next we list the appropriate sources of mass for each fermion in the model.

The new quarks (exotic) get mass from the following Yukawa couplings,
\begin{equation}
\lambda^J_{11}\bar Q_{1L}\chi J_{1R} + \lambda^J_{ij}\bar Q_{iL}\chi^* J_{jR} + h.c.
\label{exoticquarks}
\end{equation}
where $i,j=2,3$. When the $\chi$ field develops its VEV, these couplings lead to the mass matrix in the basis $(J_1\,,\,J_2\,,\,J_3)$,
\begin{equation}
M_J=
\begin{pmatrix}
\lambda^J_{11} & 0 & 0 \\
0& \lambda^J_{22} & \lambda^J_{23}\\
0 & \lambda^J_{32} & \lambda^J_{33}
\end{pmatrix}v_\chi\,
\label{massmatrixnewquarks}
\end{equation}
which, after diagonalization, leads to mass eigenvalues at $v_\chi\sim $few~TeV scale.

As for the standard quarks, their masses come from a combination of renormalizable Yukawa interactions and specific effective dimension-five operators given by,
\begin{eqnarray}
&&
 \lambda^d_{1a}\bar Q_{1L}\rho d_{aR} + \frac{\lambda^d_{ia}}{\Lambda}\varepsilon_{nmp}\left(\bar Q_{iLn}\rho_m\chi_p\right)d_{aR} + \nonumber \\
 && \lambda^u_{ia}\bar Q_{iL}\rho^* u_{aR} + \frac{\lambda^u_{1a}}{\Lambda}\varepsilon_{nmp}\left(\bar Q_{1Ln}\rho^*_m\chi^*_p\right)u_{aR} + h.c.
 \label{quarksmassterms}
\end{eqnarray}

We remember that the highest energy scale where the model is found to be perturbatively reliable is about $\Lambda=4-5 $TeV~\cite{landau}. For sake of simplicity let us assume  that $\Lambda \approx v_\chi$. In this case the up-type quarks mass matrix take the following form in the basis $(u_1 \,,\,u_2\,,\,u_3)$,
\begin{equation}
m_u \approx\frac{1}{\sqrt{2}}
\begin{pmatrix}
\lambda^u_{11}v_\rho  & \lambda^u_{12}v_\rho & \lambda^u_{13}v_\rho   \\
-\lambda^u_{21} v_\rho& -\lambda^u_{22}v_\rho &- \lambda^u_{23}v_\rho\\
-\lambda^u_{31}v_\rho & -\lambda^u_{32}v_\rho & -\lambda^u_{33}v_\rho
\end{pmatrix},
\label{massmatrixupquarks}
\end{equation}
while the mass matrix for the down-type quarks in the basis $(d_1\,,\,d_2\,,\,d_3)$ is,
\begin{equation}
m_d \approx \frac{1}{\sqrt{2}}
\begin{pmatrix}
\lambda^d_{11}v_\rho  & \lambda^d_{12}v_\rho & \lambda^d_{13}v_\rho   \\
\lambda^d_{21} v_\rho& \lambda^d_{22}v_\rho & \lambda^d_{23}v_\rho\\
\lambda^d_{31}v_\rho & \lambda^d_{32}v_\rho & \lambda^d_{33}v_\rho
\end{pmatrix}\,.
\label{massmatrixdownquarks}
\end{equation}

Finally, the charged lepton masses arise from the effective dimension five operator,
\begin{equation}
\frac{\kappa}{\Lambda}\left(\overline{f^c_L}\rho^*\right)\left(\chi^\dagger f_L \right) + h.c.
\label{chargedleptonmasses}
\end{equation}
This non-renormalizable operator generates a mass term for the charged leptons $m_l\approx \frac{1}{2}\kappa v_\rho$.

It is important to remark that, although some fermion masses have their origin through non-renormalizable operators, the standard fermions end up to develop the same mass pattern as in the SM. In other words, we have the same fine tunning in their couplings as we had in the SM in order to obtain their observed masses.

Concerning the neutrinos, we first call the attention to the fact that, whatever the particle content of the minimal 3-3-1 model,  as  the highest energy scale  of the model  is about $\Lambda=4-5 $TeV, it generally faces difficulties in generating neutrino masses at eV scale. For example, in the original version of the minimal 3-3-1 model,  effective  dimension-five operators  will lead to neutrino masses at GeV scale. In the RM331 proposed here, the lowest order effective operator that generates neutrino mass is a dimension-7 one, 
\begin{equation}
\frac{\kappa^\prime}{\Lambda^3} \epsilon_{ijk}\epsilon_{lmn}(\bar f^{C}_{Li}\rho_j \chi_k )(f_{Ll}\rho_m \chi_n)+ h.c.
\label{neutrinomasses}
\end{equation}
which is not sufficient to generate neutrino masses at eV scale, but  represents  a little improvement in the fine tuning in the coupling when compared to the  original version. 

There is another important issue to be considered in all versions of the minimal 3-3-1 model, which was first pointed out in Ref.~\cite{truly}. It concerns the proton decay. Since the model loses its perturbative character at about $\Lambda\sim 4-5$~TeV~\cite{landau}, new interactions are supposed to emerge at this new scale, implying the appearance of effective couplings at lower energies, like those leading to fermion masses in Eqs.~(\ref{quarksmassterms}), (\ref{chargedleptonmasses}) and (\ref{neutrinomasses}). Some operators involving quarks can be very dangerous considering that such a scale, $\Lambda\sim 4-5$~TeV, is not sufficient to avoid proton decay~\cite{truly}. In the RM331 the situation is exactly the same as in the minimal 3-3-1 model, where the most dangerous proton decay operator is dimension-7,
\be
\frac{C_1\epsilon_{ijk}}{\Lambda^3}\overline{(Q_{1i_L})^c} f_{1j_L} \chi_k \overline{(u_{1_R})^c} d_{1_R} + {\mbox h.c.}\,,
\label{pdecay}
\ee
with $C_1$ a dimensionless coupling and the color indexes are omitted (but contracted through the invariant and completely antisymmetric tensor), $i,\,j,\,k$ are $SU(3)_L$ indexes and the lower index $1$ means we are taking only the first family into account. This leads to proton decay by, for example, the following interaction,
\be
\frac{C_1 v_\chi}{\Lambda^3} \bar{u}_{L}^c {e_L} \bar{u}_{R}^c d_{L} + {\mbox h.c.}\,.
\label{dim7}
\ee
We choose, as in Ref.~\cite{truly} to impose a discrete $Z_2$ symmetry over the quark fields,
\[ Q_{aL}\rightarrow -Q_{aL}\,,\,\,\,\,\,\,q_{aR}\rightarrow -q_{aR}\,, \]
that guarantees the proton stability for such effective operators.

\section{Conclusions}

In this work we have shown that the  minimal 3-3-1 model can be implemented with only two scalar triplets, a curious and simplifying feature that was not observed before (at least for the minimal model~\cite{2H331}). We have seen that such shrunk scalar sector is sufficient to engender the spontaneous breaking of the $SU(3)_C \times SU(3)_L \times U(1)_N$ symmetry to the $SU(3)_C \times U(1)_{QED}$ one and generate the correct masses of all gauge bosons and fermions, including the neutrinos. 

With only two scalar triplets it is not possible to generate  mass for all fermions through renormalizable Yukawa interactions. We have to resort to effective operators. This is particularly feasible in this minimal 3-3-1 model because the cutoff energy scale of the model  is about $\Lambda=4-5 $TeV. Due to this, on choosing appropriate dimension-five effective operators, we obtain mass terms for the charged leptons and some quarks at electroweak scale.  As for the neutrinos, we got Majorana mass terms from effective dimension-seven operators that, although not leading naturally to the observed eV mass scale, represents a small improvement in the fine tuning of the scalar-neutrino coupling as compared to the original minimal 3-3-1 model.

The main difference of the RM331 from the original minimal 3-3-1 model is the amount of physical degrees of freedom in their scalar sector. While in the original model we have twenty two physical degrees of freedom in the form of scalars, in the RM331 model only four physical degrees of freedom remains in the spectrum. Namely, $h_1$ and $h_2$ and a doubly charged bilepton scalar $h^{++}$.  The neutral scalar, $h_1$, is the lightest scalar and recovers the SM Higgs interactions, while the new scalars, $h_2$ and $h^{++}$,  have masses around TeV scale. 
The advantage of this reduced spectrum is that we can easily extract 
all the scalar-gauge boson and scalar-fermion interactions without making the usual cumbersome assumptions for the couplings in the scalar potential. In this way, we turn possible a neater treatment of the phenomenology involving the scalar sector of the model, which is going to be essential for a careful analysis of Higgs search in the minimal 3-3-1 model, as well as other interesting possibilities involving the scalars as intermediate states. 

It is important to stress that although $h_1$ can reproduce all the physics of the SM Higgs, it is not exactly the standard one because it presents new interactions with the new quarks and also intermediates  flavor changing neutral processes with the standard quarks, which may also be mediated by the neutral scalar $h_2$ and the neutral gauge boson $Z^{\prime}$. Besides, the new gauge boson degrees of freedom allows some enhancement in the diphoton Higgs decay channel~\cite{H2fotons} and may give a very different contribution to the $H\rightarrow \gamma Z$ channel, since heavy degrees of freedom can possibly be important as they do not decouple in the usual sense~\cite{russos}. Finally, the  doubly charged scalar, $h^{\pm \pm}$, can be probed at the LHC through the reaction $p+p \rightarrow e^+ + e^+  + e^- + e^-$  or in a future ILC through the reaction $e^- + e^- \rightarrow h^{--}\rightarrow e^- + e^-$, a phenomenological study we wish to pursue somewhere else.

\acknowledgments
This work was supported by Conselho Nacional de Pesquisa e
Desenvolvimento Cient\'{i}fico- CNPq (JGFJr, CASP and PSRS) and Coordena\c c\~ao de Aperfei\c coamento de Pessoal de N\'{i}vel Superior - CAPES (PRDP).

\appendix
\section*{Appendix A}
In this appendix we present the neutral and charged currents with the gauge bosons of the model.
They arise from the following matter lagrangian,
\beq
{\cal L}_{cin}&=&\bar f_{a_L} i{\cal D\!\!\!\!/}\, f_{a_L} +\bar Q_{1_L}i{\cal 
D\!\!\!\!/}\,\,Q_{1_L} +\bar Q_{i_L}i{\cal D\!\!\!\!/}\,\,Q_{i_L}+\bar u_{a_R} i{\cal 
D\!\!\!\!/}\,\,u_{a_R}+
\bar d_{a_R} i{\cal D\!\!\!\!/}\,\, d_{a_R} + \bar J_{a_R} i{\cal D\!\!\!\!/}\,\, 
J_{a_R}\,,
\label{4.1}
\eeq
where $a=1,2,3$ and,
\beq
{\cal D}_\mu=\partial_\mu + ig_N 
N_L W^N_\mu+i\frac{g}{2}\vec{\lambda} \vec{W}_\mu\,,
\label{4.2}
\eeq
is the covariant derivative for any left-handed fermion triplet ($N_L$ the corresponding $U(1)_N$ quantum number)  and,
\beq
{\cal D}_\mu=\partial_\mu + ig_N N_R W^N_\mu\,,
\label{4.3}
\eeq
is the covariant derivative for the right-handed fermion singlets ($N_R$ the corresponding $U(1)_N$ quantum number).

When the covariant derivative acts on the fermion triplets we have
\beq
\frac{g}{2}\vec{\lambda}\vec{ W}_\mu=
\left (
\begin{array}{lcr}
\frac{g}{2}(W^3_\mu+\frac{1}{\sqrt{3}}W^8_\mu)  & \frac{g}{\sqrt{2}}W^+_\mu & 
\frac{g}{\sqrt{2}}V^-_\mu  \\
 \frac{g}{\sqrt{2}}W^-_\mu & \frac{g}{2}(-W^3_\mu+\frac{1}{\sqrt{3}}W^8_\mu) & 
\frac{g}{\sqrt{2}}U^{++}_\mu  \\
\frac{g}{\sqrt{2}}V^+_\mu & \frac{g}{\sqrt{2}}U^{--}_\mu & 
-g\frac{1}{\sqrt{3}}W^8_\mu
\end{array}
\right )\,.
\label{4.4} 
\eeq
while for the  anti-triplets we have,
\beq
\frac{g}{2}\vec{{\bar \lambda}} \vec{W}_\mu=
\left (
\begin{array}{lcr}
-\frac{g}{2}(W^3_\mu+\frac{1}{\sqrt{3}}W^8_\mu)  & -\frac{g}{\sqrt{2}}W^-_\mu & 
-\frac{g}{\sqrt{2}}V^+_\mu  \\
-\frac{g}{\sqrt{2}}W^+_\mu & -\frac{g}{2}(-W^3_\mu+\frac{1}{\sqrt{3}}W^8_\mu) & 
-\frac{g}{\sqrt{2}}U^{--}_\mu  \\
-\frac{g}{\sqrt{2}}V^-_\mu & -\frac{g}{\sqrt{2}}U^{++}_\mu & 
g\frac{1}{\sqrt{3}}W^8_\mu
\end{array}
\right ) ,
\label{4.5} 
\eeq
where $\bar \lambda = -\lambda^*$.

With all this in hand we are able to obtain the charged and neutral currents involving fermions and gauge bosons.
We first present the charged currents for the leptons. On considering that the charged leptons come in a diagonal basis we obtain,
\beq
{\cal L}^{CC}_{l}= \frac{g}{\sqrt{2}}\bar \nu_{a_L} \gamma^\mu 
V^l_{PMNS}e_{a_L}W^+_\mu + \frac{g}{\sqrt{2}}\overline{e^c}_{a_L}O^V\gamma^\mu 
\nu_{a_L} V^+_\mu + \frac{g}{\sqrt{2}} \overline{e^c}_{a_L} \gamma^\mu 
e_{a_L}U^{++}_\mu +h.c,
\label{4.6}
\eeq
where $a=1,2,3$ with  $V^l_{PMNS}=V^{\nu \dagger}_L $ being  the PMNS mixing  matrix and   $O^V=V^{\nu }_L$. The matrix $V^{\nu }_L$ diagonalizes the neutrino mass matrix.

For the  quarks the charged currents take the following form,
\beq
{\cal L}^{cc}_{q}&=& \frac{g}{\sqrt{2}}\bar u_L V^q_{CKM} \gamma^\mu d_LW^+_\mu 
+\frac{g}{\sqrt{2}}\left(\bar J_{1_L}\gamma^\mu (V^u_L)_{1a}u_{a_L} - \bar 
d_{l_L}(V^{d \dagger}_L)_{li}\gamma^\mu J_{i_L}\right)V^+_\mu 
\nonumber \\
&&+\frac{g}{\sqrt{2}}\left(\bar J_{1_L}\gamma^\mu (V^d_L)_{1a}d_{1_L} + \bar 
u_{l_L}(V^{u \dagger}_L)_{li}\gamma^\mu 
J_{i_L}\right)U^{++}_\mu +h.c,
\label{4.7}
\eeq
where  $a=1,2,3$, while  $i,l=2,3$. $V^q_{CKM}=V^{u \dagger}_L V^d_L$ is the Cabibbo-Kobayashi-Maskawa mixing matrix .  In the charged current above, $V^{u,d}_L$  are the mixing matrices that connect the left-handed up and down quarks symmetry eigenstates with the mass eigenstates. We assume that the new quarks come in a diagonal basis.

Next we present the neutral current among the fermions and the neutral gauge bosons $Z_\mu$  and $Z^{\prime}_\mu$.

The neutral currents for leptons are,
\beq
{\cal L}^{NC}_{l}&=&-\frac{g}{2c_W}\bar \nu_{a_L}\gamma^\mu \nu_{a_L}Z_\mu -\frac{g}{2c_W} 
\sqrt{\frac{h_W}{3}}\bar \nu_{a_L}\gamma^\mu \nu_{a_L}Z^{\prime}_\mu 
\nonumber \\
&&-\frac{g}{2c_W}\bar e_a\gamma^\mu \left(a_1 - b_1 \gamma^5 \right)e_aZ_\mu 
-\frac{g}{2c_W}\bar e_a\gamma^\mu \left(a_2 - b_2 \gamma^5 \right)e_aZ^{\prime}_\mu,
\label{4.8}
\eeq
where,
\beq
&&a_1=-\frac{1}{2} h_W \,\,\, , \,\,\,b_1 = 
-\frac{1}{2}\nonumber \\
&&a_2=\frac{1}{2}\sqrt{3h_W} \,\,\, , \,\,\,b_2 = 
-\frac{1}{2}\sqrt{3h_W}.
\label{4.9}
\eeq

The neutral currents among the up quarks  and the standard neutral gauge boson $Z_\mu$ take the form,
\beq
{\cal L}^{NC}_{Z^1 , u}&=&\frac{g}{2C_W}\left( \bar u 
\gamma_\mu[a_{1L}(u)(1-\gamma_5) +a_{1R}(u)(1+\gamma_5)]u + \right)Z_\mu^1 ,
\label{4.10} 
\eeq
where
\beq
&&a_{1L}(u)=(-\frac{1}{2}+\frac{2}{3}s^2_W) \,\,\, , \,\,\, 
a_{1R}(u)=\frac{2}{3}s^2_W.
\label{4.11}
\eeq 

The neutral currents among the  down quarks and the standard neutral gauge boson $Z_\mu$ are,
\beq
{\cal L}^{NC}_{Z^1 ,d}&=&\frac{g}{2c_W}\left( \bar d 
\gamma_\mu[a_{1L}(d)(1-\gamma_5) +a_{1R}(d)(1+\gamma_5)]d \right)Z_\mu,
\label{4.12} 
\eeq
where
\beq
&&a_{1L}(d)=(\frac{1}{2}-\frac{1}{3}s^2_W)\,\,\, , \,\,\, 
a_{1R}(d)=-\frac{1}{3}s^2_W .
\label{4.13} 
\eeq

The neutral currents among the  new (exotic) quarks and the standard neutral gauge boson $Z_\mu$ take the form,
\beq
{\cal L}^{NC}_{Z^1 ,J}&=&\frac{g}{2c_W}\left( \bar J_1 
\gamma_\mu[a_{1L}(J_1)(1-\gamma_5) +a_{1R}(J_1)(1+\gamma_5)]J_1 +\right. 
\nonumber \\
&&  \left. \bar J \gamma_\mu[a_{1L}(J)(1-\gamma_5) +a_{1R}(J)(1+\gamma_5)]J 
\right)Z_\mu^1 ,
\label{4.14} 
\eeq
where 
\beq
&&a_{1L}(J_1)=(\frac{5s^2_W}{3})\
,\,\,  \,\,\, a_{1R}(J_1)=(\frac{5s^2_W}{3} ) ,\nonumber \\
&& 
a_{1L}(J)=(-\frac{4s^2_W}{3})\,\, 
, \,\, a_{1R}(J)=(-\frac{4s^2_W}{3} ).
\label{4.15}
\eeq 

The neutral currents among the  up quarks and the new neutral gauge boson $Z^{\prime}_\mu$ take the form
\beq
{\cal L}^{NC}_{Z^{\prime} , u}&=&\frac{g}{2c_W}\left( \bar u 
\gamma_\mu[a_{2R}(u)(1+\gamma_5)]u +\right. \nonumber \\
&& \left. \bar u \gamma_\mu V^{u *}_L Y_{Z^{\prime}}^{um}V^u_L(1-\gamma_5) u  
\right)Z_\mu^2 ,
\label{4.16} 
\eeq
where
\beq
&&a_{1R}(u)=-\frac{2}{\sqrt{3h_W}}s^2_W \,\,\,\,,\,\,\,\, Y_{Z^{\prime}}^{um}=\frac{1}{\sqrt{12h_W}}\mbox{diag}(1,1-2s^2_W , 1-2s^2_W ).
\label{4.17}
\eeq 

The neutral currents among the  down quarks and the new neutral gauge boson $Z^{\prime}_\mu$ take the form
\beq
{\cal L}^{NC}_{Z^{\prime} ,d}&=&\frac{g}{2c_W}\left( \bar d 
\gamma_\mu[a_{2R}(d)(1+\gamma_5)]d +\right. \nonumber \\
&& \left. \bar d \gamma_\mu V^{d *}_L Y_{Z^2}^{dm}V^d_L(1-\gamma_5) d 
\right)Z_\mu^2 ,
\label{4.16} 
\eeq
where 
\beq
&&a_{2R}(d)=\frac{1}{\sqrt{3h_W}}s^2_W \,\,\,\,,\,\,\,\,Y_{Z^{\prime}}^{md}=\frac{1}{\sqrt{12h_W}}diag(1,1-2s^2_W , 1-2s^2_W ) .
\label{4.17}
\eeq

The neutral currents among the  new (exotic) quarks and the new neutral gauge boson $Z^{\prime}_\mu$ take the form
\beq
{\cal L}^{NC}_{Z^{\prime} ,J}&=&\frac{g}{2c_W}\left( \bar J_1 
\gamma_\mu[a_{2L}(J_1)(1-\gamma_5) +a_{2R}(J_1)(1+\gamma_5)]J_1 +\right. 
\nonumber \\
&&  \left. \bar J \gamma_\mu[a_{2L}(J)(1-\gamma_5) +a_{2R}(J)(1+\gamma_5)]J 
\right)Z_\mu^2 ,
\label{4.18} 
\eeq
where
\beq 
&&a_{2L}(J_1)=\frac{(1-9s^2_W)}{\sqrt{3h_W}}\,\, , \,\, a_{2R}(J_1)=- 
\frac{5s^2_W}{\sqrt{3h_W}} ,\nonumber \\
&& 
a_{2L}(J)=-\frac{(1-5s^2_W)}{\sqrt{3h_W}}\,\, 
, \,\,  a_{2R}(J)=+ 
\frac{4s^2_W}{\sqrt{3h_W}}.
\label{4.19}
\eeq 

$h_W=1-4s^2_W$.

\section*{Appendix B}
In this appendix we present all the trilinear and quartic interactions among the gauge bosons and the scalars of the model. All these terms arise from the Lagrangian,
\begin{equation}
\mathcal{L=}\left( \mathcal{D}_{\mu }\chi \right) ^{\dagger }\left( \mathcal{%
D}^{\mu }\chi \right) +\left( \mathcal{D}_{\mu }\rho \right) ^{\dagger
}\left( \mathcal{D}^{\mu }\rho \right) .
\label{covascalar}
\end{equation}%

First we present the trilinear terms which are composed by the following interactions
\begin{eqnarray}
\begin{tabular}{|l|l|}
\hline
Vertex  & Coupling \\ \hline
$W^{+}W^{-}h^{1}$ & $\frac{1}{2}c_{\beta }g^{2}v_{\rho }$ \\ \hline
$V^{+}V^{-}h^{1}$ & $-\frac{1}{2}s_{\beta }g^{2}v_{\chi }$ \\ \hline
$U^{++}U^{--}h^{1}$ & $\frac{1}{2}\left( c_{\beta }v_{\rho }-s_{\beta
}v_{\chi }\right) g^{2}$ \\ \hline
$ZZh_{1}$ & $\frac{1}{4}g^{2}v_{\rho }c_{\beta }c_{\theta }^{2}\sec
^{2}_W$ \\ \hline
$Z^{\prime}Z^{\prime}h_{1}$ & $\frac{g^{2}c_{\theta }^{2}}{12h_W}\left[ h_W^{2}\left(
c_{\beta }v_{\rho }-4s_{\beta }v_{\chi }\right) \sec ^{2}_W+6k_{1}t_W^2%
\right] $ \\ \hline
$ZZ^{\prime}h_{1}$ & $-\frac{1}{2\sqrt{3}\sqrt{h_W}}g^{2}c_{\beta }\left(
3+h_W-3c_{2W}\right) c_{\theta }^{2}\sec ^{2}_W$ \\ \hline
$W^{+}W^{-}h^{2}$ & $\frac{1}{2}s_{\beta }g^{2}v_{\rho }$ \\ \hline
$V^{+}V^{-}h^{2}$ & $\frac{1}{2}c_{\beta }g^{2}v_{\chi }$ \\ \hline
$U^{++}U^{--}h^{2}$ & $\frac{1}{2}\left( s_{\beta }v_{\rho }+c_{\beta
}v_{\chi }\right) g^{2}$ \\ \hline
$ZZh_{2}$ & $\frac{1}{4}g^{2}v_{\rho }s_{\beta }c_{\theta }^{2}\sec
^{2}_W$ \\ \hline
$ZZh_{2}$ & $\frac{g^{2}c_{\theta }^{2}}{12h_W}[h_W^{2}\left( s_{\beta
}v_{\rho }+4c_{\beta }v_{\chi }\right) \sec ^{2}_W+12k_{2}t_W^2]$ \\ 
\hline
$Z^{\prime}Z^{\prime}h_{2}$ & $-\frac{1}{2\sqrt{3}\sqrt{h_W}}g^{2}s_{\beta }\left(
3+h_W-3c_{2W}\right) c_{\theta }^{2}\sec ^{2}_W$ \\ \hline
$W^{\pm }V^{\pm }h^{\pm \pm }$ & $\frac{1}{2\sqrt{2}}\left( c_{\alpha
}v_{\rho }+s_{\alpha }v_{\chi }\right) g^{2}$ \\ \hline
$U^{\pm \pm }Ah^{\mp \mp }$ & $g^{2}\left( c_{\alpha }v_{\rho }-s_{\alpha
}v_{\chi }\right) s_{W}$ \\ \hline
$U^{\pm \pm }Zh^{\mp \mp }$ & $\frac{1}{4}g^{2}c_{\theta }\left[
-3c_{\alpha }v_{\rho }+s_{\alpha }v_{\chi }+2\left( c_{\alpha }v_{\rho
}-s_{\alpha }v_{\chi }\right) c_{2W}\right] \sec_W$ \\ \hline
$A_{\mu }h^{++}h^{--}$ & $-\frac{3}{2}igs_{W}\left( k_{1}-k_{2}\right) _{\mu
}$ \\ \hline
$Z_{\mu }h^{++}h^{--}$ & $-\frac{1}{4}igc_{\theta }\left[ -3c_{\alpha
}^{2}-s_{\alpha }^{2}+3c_{2W}\right] \sec_W\left( k_{1}-k_{2}\right) _{\mu }$
\\ \hline
$U_{\mu }^{\pm \pm }h_{1}h^{\mp \mp }$ & $\pm ig\cos \left( \alpha +\beta
\right) \left( k_{1}-k_{2}\right) _{\mu }$ \\ \hline
$U_{\mu }^{\pm \pm }h_{2}h^{\mp \mp }$ & $\pm ig\sin \left( \alpha +\beta
\right) \left( k_{1}-k_{2}\right) _{\mu }$ \\ \hline
\end{tabular}
\label{trilinearscalars}
\end{eqnarray}
where 
\[
k_{1}=\left( 3+2h_W\right) c_{\beta }v_{\rho }-\left( 3+4h_W\right)
s_{\beta }v_{\chi }-3\left( c_{\beta }v_{\rho }-s_{\beta }v_{\chi }\right)
c_{2W}
\]
and
\[
k_{2}=\left( s_{\beta }v_{\rho }+2c_{\beta }v_{\rho }\right)
h_W+3\left( s_{\beta }v_{\rho }+c_{\beta }v_{\chi }\right) s_{W}^{2} .
\]

The quartic terms are composed by the following interactions, 
\begin{equation}
\begin{tabular}{|l|l|}
\hline
Vertex & Coupling \\ \hline
$W^{+}W^{-}h^{1}h^{1}$ & $\frac{1}{4}c_{\beta }^{2}g^{2}$ \\ \hline
$V^{+}V^{-}h^{1}h^{1}$ & $\frac{1}{4}s_{\beta }^{2}g^{2}$ \\ \hline
$U^{++}U^{--}h^{1}h^{1}$ & $\frac{1}{4}g^{2}$ \\ \hline
$ZZh_{1}h_{1}$ & $\frac{1}{8}g^{2}c_{\beta }^{2}c_{\theta }^{2}\sec
^{2}_W$ \\ \hline
$Z^{\prime}Z^{\prime}h_{1}h_{1}$ & $\frac{1}{24h_W}g^{2}c_{\theta }^{2}\left[
k_{3}+12\left( h_W\left( c_{\beta }^{2}+2s_{\beta }^{2}\right)
+3s_{W}^{2}\right) t_W^2\right] $ \\ \hline
$ZZ^{\prime}h_{1}h_{1}$ & $-\frac{1}{4\sqrt{3}\sqrt{h_W}}g^{2}c_{\beta
}^{2}\left( 3+h_W-3c_{2W}\right) c_{\theta }^{2}\sec ^{2}_W$ \\ \hline
$W^{+}W^{-}h^{2}h^{2}$ & $\frac{1}{4}s_{\beta }^{2}g^{2}$ \\ \hline
$V^{+}V^{-}h^{2}h^{2}$ & $\frac{1}{4}c_{\beta }^{2}g^{2}$ \\ \hline
$U^{++}U^{--}h^{2}h^{2}$ & $\frac{1}{4}g^{2}$ \\ \hline
$ZZh_{2}h_{2}$ & $\frac{1}{8}g^{2}s_{\beta }^{2}c_{\theta }^{2}\sec
^{2}_W$ \\ \hline
$Z^{\prime}Z^{\prime}h_{2}h_{2}$ & $\frac{1}{24h_W}g^{2}c_{\theta }^{2}\left[
k_{4}+12\left( h_W\left( 2c_{\beta }^{2}+s_{\beta }^{2}\right)
+3s_{W}^{2}\right) t_W^2\right] $ \\ \hline
$ZZ^{\prime}h_{2}h_{2}$ & $-\frac{1}{12h_W}g^{2}s_{\beta }^{2}c_{\theta
}\left( \sqrt{3}\left( 3+h_W-3c_{2W}\right) c_{\theta }^{2}\sec ^{2}_W-6\sqrt{%
h_W}s_{W}^{2}s_{\theta }\right) $ \\ \hline
$W^{+}W^{-}h^{1}h^{2}$ & $\frac{1}{2}c_{\beta }s_{\beta }g^{2}$ \\ \hline
$V^{+}V^{-}h^{1}h^{2}$ & $-\frac{1}{2}c_{\beta }s_{\beta }g^{2}$ \\ \hline
$ZZh_{1}h_{2}$ & $\frac{1}{8}g^{2}s_{2\beta }c_{\theta }^{2}\sec
^{2}_W$ \\ \hline
$Z^{\prime}Z^{\prime}h_{1}h_{2}$ & $-\frac{1}{8}g^{2}s_{2\beta }c_{\theta }^{2}\left[
h_W\sec ^{2}_W+4t_W^2\right] $ \\ \hline
$ZZ^{\prime}h_{1}h_{2}$ & $-\frac{1}{4\sqrt{3}\sqrt{h_W}}g^{2}c_{\theta
}^{2}\left( -6+h_W-3c_{2W}\right) c_{\theta }^{2}\sec ^{2}_W$ \\ \hline
$W^{+}W^{-}h^{++}h^{--}$ & $\frac{1}{2}s_{\alpha }^{2}g^{2}$ \\ \hline
$V^{+}V^{-}h^{++}h^{--}$ & $\frac{1}{2}c_{\alpha }^{2}g^{2}$ \\ \hline
$ZZh^{++}h^{--}$ & $\frac{1}{4}g^{2}\left[ 3\left( 2c_{\alpha
}^{2}-s_{\alpha }^{2}\right) -4\left( 2c_{\alpha }^{2}+s_{\alpha
}^{2}\right) c_{2W}+2c_{2W}\right] c_{\theta }^{2}\sec ^{2}_W$ \\ \hline
$Z^{\prime}Z^{\prime}h^{++}h^{--}$ & $\frac{g^{2}c_{\theta }^{2}}{12h_W}\left[
k_{5}+h_W^{2}\left( 4c_{\alpha }^{2}+s_{\alpha }^{2}\right) \sec ^{2}_W \right] 
$ \\ \hline
$ZZ^{\prime}h^{++}h^{--}$ & $\frac{1}{2\sqrt{3}h_W}g^{2}c_{\theta }^{2}\left[
k_{6}+h_W\left( -s_{\alpha }^{2}+\left( 8c_{\alpha }^{2}+3s_{\alpha
}^{2}\right) t_W^2\right) \right] $ \\ \hline
$U^{++}U^{--}h^{++}h^{--}$ & $\frac{1}{2}g^{2}$ \\ \hline
$AAh^{++}h^{--}$ & $4g^{2}s_{W}^{2}$ \\ \hline
$W^{\pm }V^{\pm }h^{1}h^{\mp \mp }$ & $\frac{1}{2\sqrt{2}}g^{2}\cos \left(
\alpha +\beta \right) $ \\ \hline
$U^{\pm \pm }Ah^{1}h^{\mp \mp }$ & $g^{2}s_{w}^{2}\cos \left( \alpha -\beta
\right) $ \\ \hline
$W^{\pm }V^{\pm }h^{2}h^{\mp \mp }$ & $\frac{1}{2\sqrt{2}}g^{2}\sin \left(
\alpha +\beta \right) $ \\ \hline
$U^{\pm \pm }Ah^{\mp \mp }h^{2}$ & $g^{2}s_{w}^{2}\sin \left( \beta -\alpha
\right) $ \\ \hline
\end{tabular}%
\end{equation}
where
\[ 
k_{3}=h_W^{2}\left( c_{\beta }^{2}+4s_{\beta }^{2}\right) \sec ^{2}_W ,
\]
\[
k_{4}=h_W^{2}\left( 4c_{\beta }^{2}+s_{\beta }^{2}\right) \sec ^{2}_W ,
\]
\[
k_{5}=-6\left( c_{\alpha }^{2}\left( -3+4h_W\right) +\left(
-3+2h_W\right) s_{\alpha }^{2}+3c_{2W}\right) t_W^2 
\]
and
\[
k_{6}=-6s_{W}^{2}\left( -s_{\alpha }^{2}+\left( 4c_{\alpha }^{2}+3s_{\alpha
}^{2}\right) t_W^2\right).
\]

\section*{Appendix C}
In this appendix we present the interactions among charged fermions and scalars. We start with the leptons. On opening the effective dimension-five operator in Eq.~(\ref{chargedleptonmasses}), we obtain the following interactions,
\begin{eqnarray}
{\cal L}_{l}&=&\dfrac{\kappa}{2}\left( \dfrac{v_\rho}{v_{\chi}}\cos(\beta) - \sin(\beta) \right)\bar{l}lh_1+\dfrac{\kappa}{2}\left( \dfrac{v_\rho}{v_{\chi}}\sin(\beta) + \cos(\beta) \right)\bar{l}lh_2 \nonumber \\ & &\mbox{}
+\dfrac{\kappa v_{\rho}\sin(\alpha)}{\sqrt{2} v_\chi}h^{--}\overline{l_L}(l^c)_R +\dfrac{\kappa \cos(\alpha)}{\sqrt{2}}h^{--}\overline{l_R}(l^c)_L  
\nonumber \\ & &\mbox{}
 -\dfrac{\kappa}{2 v_\chi}\sin(\beta)\cos(\beta)\bar{l}lh_1h_1 + \dfrac{\kappa}{2 v_\chi}\sin(\beta)\cos(\beta)\bar{l}lh_2h_2 
\nonumber \\ & &\mbox{}
+\dfrac{\kappa\sin(\alpha)\cos(\alpha)}{v_\chi}\bar{l}lh^{++}h^{--} +\dfrac{\kappa}{2 v_\chi}\cos(2\beta)\bar{l}lh_1h_2 
\nonumber \\ & &\mbox{}
+\dfrac{\kappa}{v_\chi}\left(\dfrac{-\sin(\alpha)\sin(\beta)}{\sqrt{2}}h^{--}h_1\overline{l_L}(l^c)_R +\dfrac{\sin(\alpha)\cos(\beta)}{\sqrt{2}}h^{--}h_2\overline{l_L}(l^c)_R \right) 
\nonumber \\ & &\mbox{}
+\dfrac{\kappa}{v_\chi}\left(\dfrac{\cos(\alpha)\cos(\beta)}{\sqrt{2}}h^{--}h_1\overline{l_R}(l^c)_L +\dfrac{\cos(\alpha)\sin(\beta)}{\sqrt{2}}h^{--}h_2\overline{l_R}(l^c)_L \right)+h.c.\,,
\label{leptonsscalarinteractions}
\end{eqnarray}
where $l=e, \mu , \tau$.

The  renormalizable interactions among standard quarks and scalars are composed by the terms,
\begin{eqnarray}
{\cal L}&= & \bar u_{L} \Gamma^u_1 u_{R}h_1 + \bar u_{L} \Gamma^u_2 u_{R}h_2
+\bar d_{L} \Gamma^d_1 d_{R}h_1 +\bar d_{L} \Gamma^d_2 d_{R}h_2+h.c,
\label{quarksscalarinteractions} 
\end{eqnarray}
where $u=\left( u\,,\,c\,,\,t  \right)$ and $d=\left( d\,,\,s\,,\,b \right)$ and 
\begin{eqnarray}
\Gamma_1^u = \dfrac{\sin(\beta)}{\sqrt{2}} \left(\begin{matrix}
\dfrac{\lambda^u_{11}}{\sqrt{2}}\left(\dfrac{v_\rho \cos(\beta)}{v_\chi \sin(\beta)} - 1\right)  & \dfrac{\lambda^u_{12}}{\sqrt{2}}\left(\dfrac{v_\rho \cos(\beta)}{v_\chi \sin(\beta)} - 1\right) & \dfrac{\lambda^u_{13}}{\sqrt{2}}\left(\dfrac{v_\rho \cos(\beta)}{v_\chi \sin(\beta)} - 1\right) \\ 
\lambda^u_{21} & \lambda^u_{22} & \lambda^u_{23} \\ 
\lambda^u_{31} & \lambda^u_{32} & \lambda^u_{33}
\end{matrix}\right);  
\label{G1}
\end{eqnarray}

\begin{eqnarray}
\Gamma_2^u = \dfrac{\cos(\beta)}{\sqrt{2}} \left(\begin{matrix}
\dfrac{\lambda^u_{11}}{\sqrt{2}}\left(\dfrac{v_\rho \sin(\beta)}{v_\chi \cos(\beta)} + 1\right)  & \dfrac{\lambda^u_{12}}{\sqrt{2}}\left(\dfrac{v_\rho \sin(\beta)}{v_\chi \cos(\beta)} + 1\right) & \dfrac{\lambda^u_{13}}{\sqrt{2}}\left(\dfrac{v_\rho \sin(\beta)}{v_\chi \cos(\beta)} + 1\right) \\ 
-\lambda^u_{21} & -\lambda^u_{22} & -\lambda^u_{23} \\ 
-\lambda^u_{31} & -\lambda^u_{32} & -\lambda^u_{33}
\end{matrix}\right);  
\label{G2}
\end{eqnarray}

\begin{eqnarray}
\Gamma_1^d = \dfrac{\sin(\beta)}{\sqrt{2}} \left(\begin{matrix}
-\lambda^d_{11} & -\lambda^d_{12} & -\lambda^d_{13} \\ 
\dfrac{\lambda^d_{21}}{\sqrt{2}}\left(\dfrac{v_\rho \cos(\beta)}{v_\chi \sin(\beta)} - 1\right)  & \dfrac{\lambda^d_{22}}{\sqrt{2}}\left(\dfrac{v_\rho \cos(\beta)}{v_\chi \sin(\beta)} - 1\right) & \dfrac{\lambda^d_{23}}{\sqrt{2}}\left(\dfrac{v_\rho \cos(\beta)}{v_\chi \sin(\beta)} - 1\right) \\ 
\dfrac{\lambda^d_{31}}{\sqrt{2}}\left(\dfrac{v_\rho \cos(\beta)}{v_\chi \sin(\beta)} - 1\right)  & \dfrac{\lambda^d_{32}}{\sqrt{2}}\left(\dfrac{v_\rho \cos(\beta)}{v_\chi \sin(\beta)} - 1\right) & \dfrac{\lambda^d_{33}}{\sqrt{2}}\left(\dfrac{v_\rho \cos(\beta)}{v_\chi \sin(\beta)} - 1\right) 
\end{matrix}\right);  
\label{G3}
\end{eqnarray}

\begin{eqnarray}
\Gamma_2^d = \dfrac{\cos(\beta)}{\sqrt{2}} \left(\begin{matrix}
\lambda^d_{11} & \lambda^d_{12} & \lambda^d_{13} \\ 
\dfrac{\lambda^d_{21}}{\sqrt{2}}\left(\dfrac{v_\rho \sin(\beta)}{v_\chi \cos(\beta)} + 1\right)  & \dfrac{\lambda^d_{22}}{\sqrt{2}}\left(\dfrac{v_\rho \sin(\beta)}{v_\chi \cos(\beta)} + 1\right) & \dfrac{\lambda^d_{23}}{\sqrt{2}}\left(\dfrac{v_\rho \sin(\beta)}{v_\chi \cos(\beta)} + 1\right)\\ 
\dfrac{\lambda^d_{31}}{\sqrt{2}}\left(\dfrac{v_\rho \sin(\beta)}{v_\chi \cos(\beta} + 1\right)  & \dfrac{\lambda^d_{32}}{\sqrt{2}}\left(\dfrac{v_\rho \sin(\beta)}{v_\chi \cos(\beta)} + 1\right) & \dfrac{\lambda^d_{33}}{\sqrt{2}}\left(\dfrac{v_\rho \sin(\beta)}{v_\chi \cos(\beta)} + 1\right) 
\end{matrix}\right). 
\label{G4}
\end{eqnarray}



\end{document}